\documentclass{article}

\PassOptionsToPackage{numbers, compress}{natbib}
\usepackage{natbib}

\usepackage[final]{tackling_climate_workshop_style}

\usepackage[utf8]{inputenc} 
\usepackage[T1]{fontenc}    
\usepackage{hyperref}       
\usepackage{url}            
\usepackage{booktabs}       
\usepackage{amsfonts}       
\usepackage{nicefrac}       
\usepackage{microtype}      
\usepackage{comment}
\usepackage{xcolor}

\usepackage{makecell}
\usepackage{graphicx} 
\usepackage{caption}  
\usepackage{booktabs}
\usepackage{titlesec}

\title{Quantifying Climate Policy Action and Its Links to Development Outcomes: A Cross-National Data-Driven Analysis}

\author{%
  Aditi Dutta\\
  School of Government\\
  University of Birmingham\\
  United Kingdom\\
  \texttt{a.dutta.4@bham.ac.uk} \\
}

\begin{document}

\maketitle

\begin{abstract}
Addressing climate change effectively requires more than cataloguing the number of policies in place; thus it calls for tools that can predict their themes or subject, and analyze their tangible impacts on development outcomes. Existing assessments often rely on qualitative descriptions or composite indices, which can mask crucial differences between key domains such as mitigation, adaptation, disaster risk management, and loss and damage. To bridge this gap, we develop a quantitative indicator of climate policy orientation by applying a multilingual transformer-based language model to official national policy documents, achieving a classification accuracy of 0.90 (F1-score). Linking these indicators with World Bank development data in panel regressions reveals that mitigation policies are associated with higher GDP and GNI; disaster risk management correlates with greater GNI and debt but reduced foreign direct investment; adaptation and loss and damage show limited measurable effects. This integrated NLP–econometric framework enables comparable, theme-specific analysis of climate governance, offering a scalable method to monitor progress, evaluate trade-offs, and align policy emphasis with development goals. The code and datasets used in this study are publicly available at: \url{https://github.com/booktrackerGirl/climate_change_policy_analysis}.

\end{abstract}

\section{Introduction}

Assessments of the impacts of climate change have shifted from academic inquiry toward operational, stakeholder-driven approaches \cite{change2014impacts, mechler2016identifying}. The United Nations has played a key role in advancing the post-2015 development agenda, fostering global consultations, and supporting Member States through evidence-based inputs, analytical guidance, and field expertise \cite{intergovernmental}.
The implementation of climate policy action plans is in the shared interest of all stakeholders—particularly across developing countries seeking to achieve growth, health, and development objectives with direct or indirect climate co-benefits. Quantitative indicators can strengthen such processes by supporting evidence-based governance \citep{cust2009using}. While environmental health indicators have long helped monitor complex trends and inform policy \citep{liu2021toward}, major research gaps persist due to interacting drivers and uncertainties in linking climate and health outcomes \citep{haines2019imperative, tong2019preventing, hambling2011review}. Integrating environmental and epidemiological data can improve understanding of how climate drivers shape environmental states, human exposure, and ultimately health outcomes \cite{hambling2011review, briggs1999environmental, kjellstrom1995framework, liu2021toward}.
Beyond these sectoral insights, the Paris Agreement mandates countries to implement and report on their adaptation progress \cite{magnan2016global, lesnikowski2017does, berrang2019tracking}. As the volume of national climate laws and policies expands, text-based analysis has become an important tool for synthesizing information on governance and national engagement \cite{climatepolicytrackerpipeline}, enabling systematic policy comparison and future planning. Yet, existing policy-tracking efforts remain largely descriptive, lacking a quantitative, theme-specific framework that captures differences in emphasis across Mitigation, Adaptation, Disaster Risk Management, and Loss and Damage, and links these to measurable development outcomes.
This study addresses that gap by developing a replicable, text-based indicator of national climate policy emphasis using machine learning, and empirically examining its association with socioeconomic outcomes. The resulting framework supports more granular cross-country comparison and evidence-based climate governance, contributing to emerging efforts to systematically quantify policy ambition and alignment with global development goals.

\section{Methodology}

The methodology follows two steps: (1) Classification Task, and (2) Statistical Analysis.

Step 1 (Climate Policy Classification): We develop a quantitative text-based indicator of national climate policy ambition using official policies from the Climate Change Laws of the World (CCLW) database \citep{ccl}. The dataset contains climate law and policy summaries labeled with thematic tags—Adaptation, Mitigation, Disaster Risk Management, or Loss and Damage. A supervised multi-label DistilBERT model \citep{Sanh2019DistilBERTAD} is fine-tuned on these summaries to generate dense text embeddings and assign themes automatically, without relying on hand-crafted features or external metadata. This enables standardized, comparable indicators of policy orientation across languages and countries.

Step 2 (Statistical Analysis): We link these theme-specific indicators to socioeconomic outcomes from the World Bank World Development Indicators (WDI) \citep{worldbank_wdi} for 2015 onward. The analysis proceeds in three stages: (1) Descriptive analysis through faceted boxplots identifies leading countries for each policy theme; (2) Correspondence analysis (CA) uncovers latent relationships between countries and policy areas; and (3) Two-way fixed-effects panel regressions estimate associations between each policy theme and key development indicators such as GDP, GNI, FDI, debt stocks, and electricity consumption.

Together, these steps form a coherent, text-based cross-national framework: the NLP model quantifies policy emphasis from unstructured texts, and the econometric analysis evaluates how thematic priorities align with national development patterns, offering a scalable approach to evidence-based climate policy tracking.

\section{Experiment and Results}

\subsection{Classification Task}
\begin{minipage}[b]{0.5\textwidth}
\centering
\tiny
\begin{tabular}{lcccc}
\toprule
\textbf{Category} & \textbf{Precision} & \textbf{Recall} & \textbf{F1-Score} & \textbf{Support} \\
\midrule
\makecell{Adaptation} & 0.82 & 0.87 & 0.84 & 247 \\
\makecell{Disaster Risk \\Management} & 0.77 & 0.66 & 0.71 & 83  \\
\makecell{Loss and Damage} & 1.00 & 0.36 & 0.53 & 11  \\
\makecell{Mitigation} & 0.95 & 0.97 & 0.96 & 498 \\
\midrule
Micro Avg                 & 0.90 & 0.90 & 0.90 & 839 \\
Macro Avg                 & 0.89 & 0.72 & 0.76 & 839 \\
Weighted Avg              & 0.90 & 0.90 & 0.90 & 839 \\
Samples Avg               & 0.92 & 0.93 & 0.91 & 839 \\
\bottomrule
\end{tabular}
\captionof{table}{\small This table shows the classification report (threshold = 0.5) for the multi-label DistilBERT predicting policy themes from climate policy summaries.}
\label{tab:classification_report}
\end{minipage}
\hfill
\begin{minipage}[b]{0.45\textwidth}
  \centering
  \includegraphics[width=0.7\linewidth]{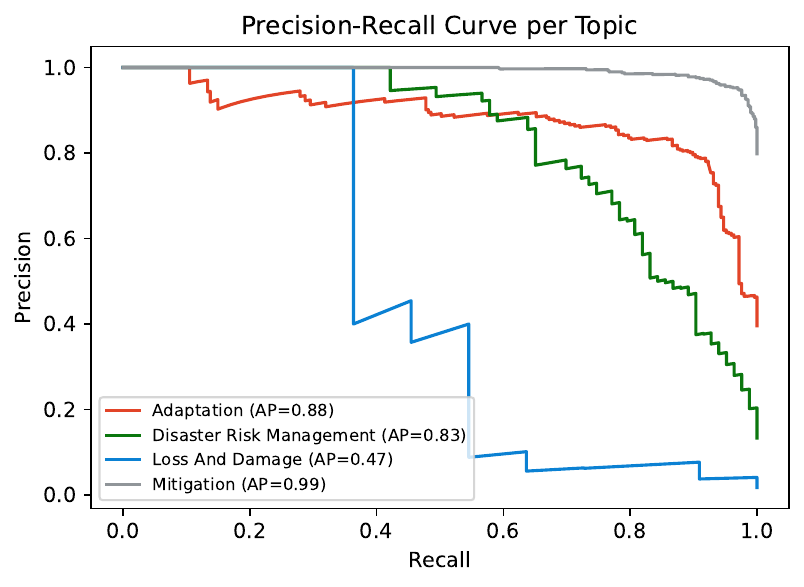}
  \vspace{-5pt}
  \captionof{figure}{\small This figure shows a precision-recall curve for the multi-label classifier. The area under the curve (average precision, AP) provides a single-number summary.}
  \label{fig:precision_recall}
\end{minipage}

The precision–recall curve (Figure \ref{fig:precision_recall}) and classification report (Table \ref{tab:classification_report}) summarize model performance. Results closely follow class frequency: Mitigation (498 samples) performs best (AP = 0.99, 95\% precision, 97\% recall), while Loss \& Damage (11 samples) performs worst (AP = 0.47, 100\% precision, 36\% recall) due to extreme class imbalance. Adaptation shows balanced results (AP = 0.88, 82\% precision, 87\% recall), and Disaster Risk Management achieves moderate discrimination (AP = 0.83) but lower recall (66\%).
Overall micro/macro F1 = 0.90, though category disparities remain masked by the dominance of Mitigation. The model's tendency toward high precision but lower recall for underrepresented categories suggests it has learned to be conservative when uncertain, prioritizing avoiding false positives over capturing all relevant instances.


\subsection{Statistical Analysis}

Descriptive summary statistics (Figure \ref{fig:boxplot}) show expected cross-country variation: developed economies allocate greater emphasis to mitigation, while climate-vulnerable small island states prioritize adaptation and disaster risk management, with Loss \& Damage remaining limited globally.
To further characterize cross-country variation in thematic climate policy emphasis, we apply correspondence analysis (Figure \ref{fig:correspondence_plot}) supports these patterns, with two dimensions explaining 92.1\% of variance. This allows us to visualize how countries cluster based on their relative prioritization of mitigation, adaptation, disaster risk management, and loss and damage. Dimension 1 (71.7\%) separates developed countries with balanced policy portfolios from SIDS and developing states with narrower focus; Dimension 2 (20.4\%) differentiates specialization, linking Loss \& Damage to small islands (e.g., Tuvalu, Seychelles) and DRM to developing countries (e.g., Somalia, Jamaica, Bolivia). Overall, climate policy emphasis aligns closely with both resource capacity and climate risk profiles. We use correspondence analysis descriptively; deeper causal interpretation remains limited, as associations are correlational rather than causal.

\begin{minipage}[t]{0.55\textwidth}
  \centering
  \includegraphics[width=0.95\textwidth]{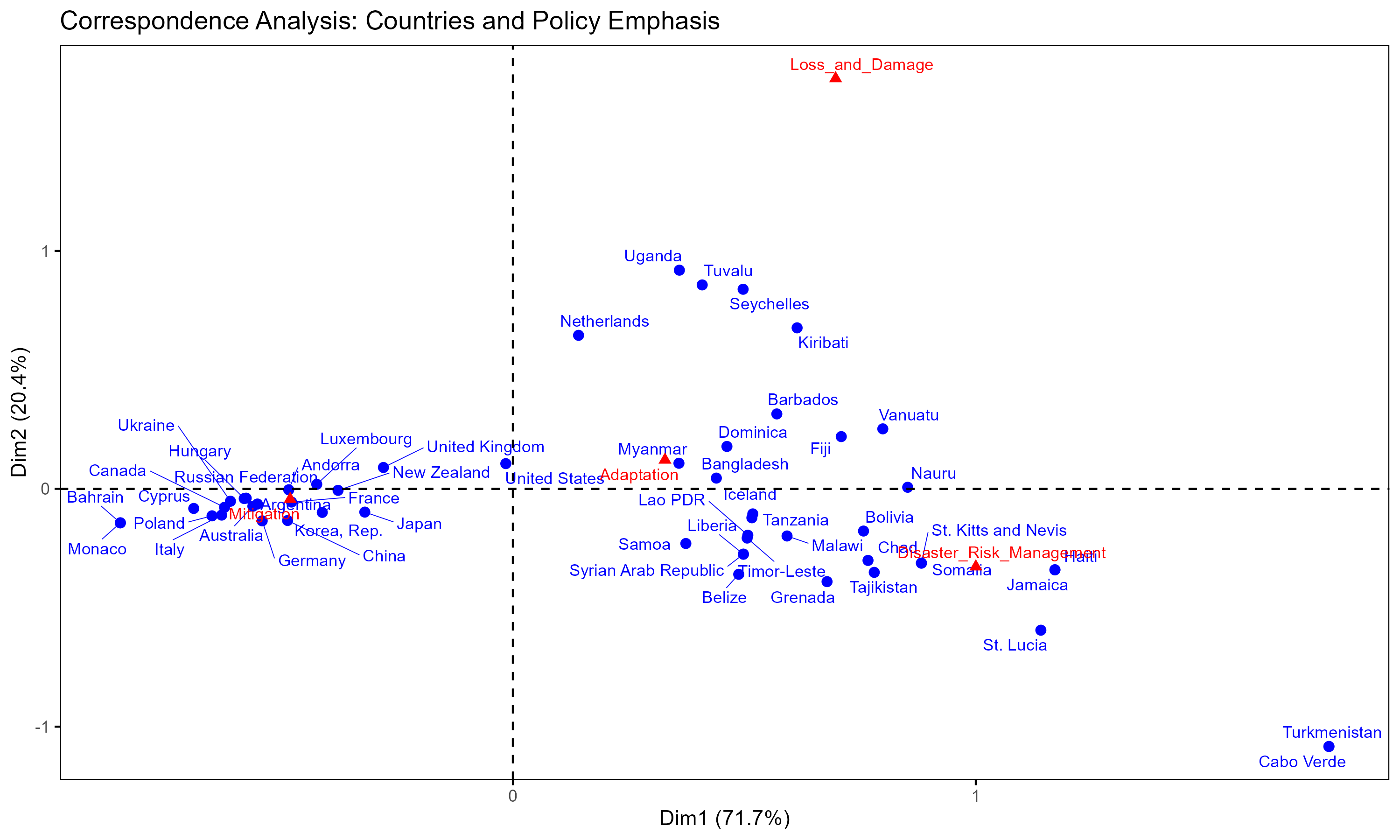}
  \captionof{figure}{\small Biplot of top 50 (+G7) countries (\textcolor{blue}{blue}) and four climate policy areas (\textcolor{red}{red}). Proximity indicates emphasis: small island states favor Adaptation/Disaster Risk Management, wealthier nations favor Mitigation.}
  \label{fig:correspondence_plot}
\end{minipage}
\hfill
\begin{minipage}[t]{0.42\textwidth}
  \centering
  \includegraphics[width=0.98\linewidth]{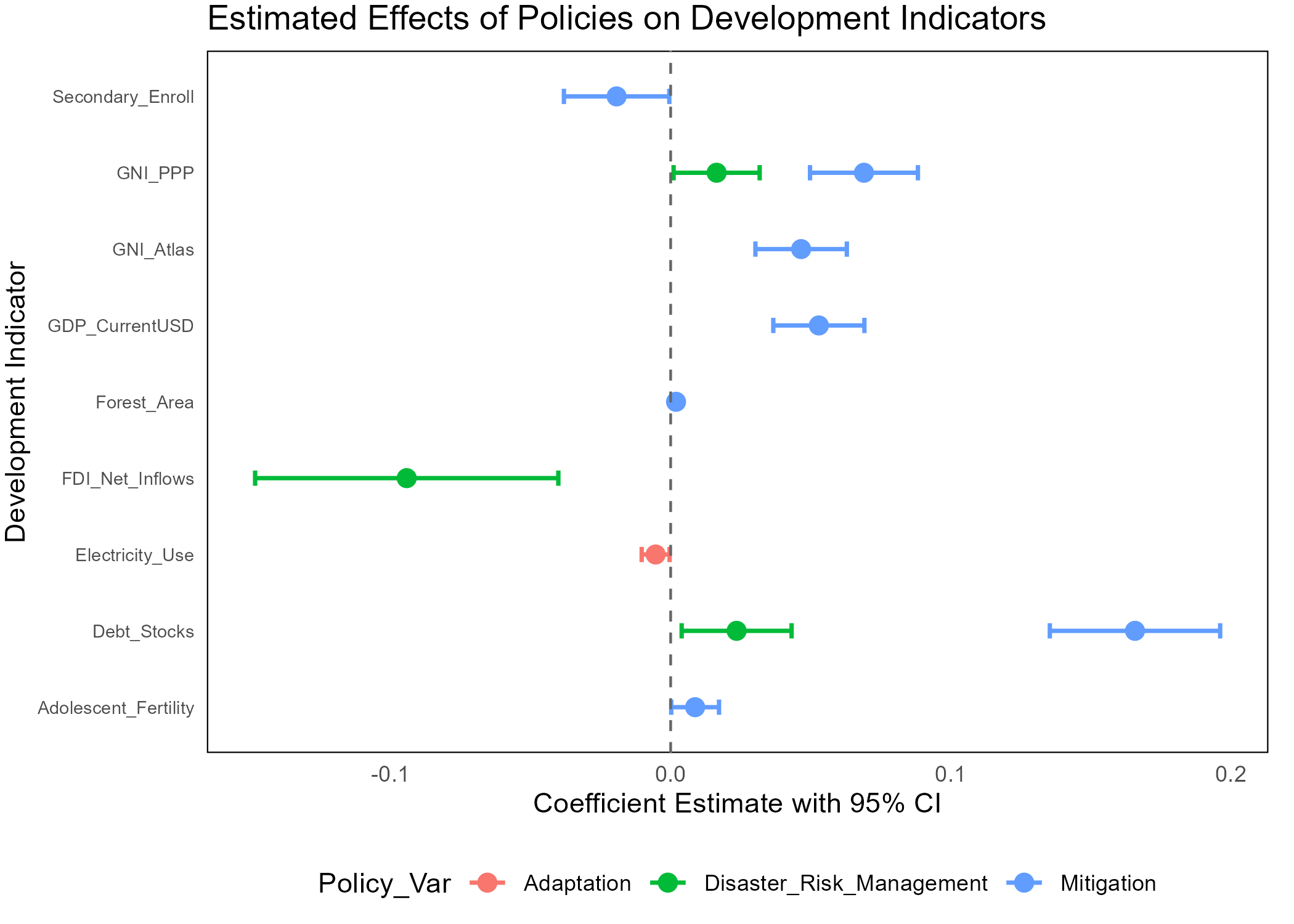}
  \captionof{figure}{\small This figure shows the estimated effects of climate policy types on development indicators with 95\% confidence intervals.}
  \label{fig:effect_plot}
\end{minipage}

Links between policy emphasis and development outcomes were examined via two-way fixed-effects panel regressions using CCLW and WDI country-year data. These results are associative rather than causal and serve as an initial foundation for future causal analysis of climate–development linkages. Four policy variables—Mitigation, Adaptation, DRM, and Loss \& Damage—were modeled jointly to capture real-world overlap in climate actions. Mitigation policies show consistently strong, significant positive effects on GDP, GNI (Atlas \& PPP), and debt stocks, suggesting growth reinforcement through climate action. This aligns with World Bank findings that a 10\% rise in GDP per capita can reduce climate risk exposure by $\approx$100 million people \citep{group_2024_12} and with evidence from the OECD that climate‑mitigation policies can yield long‑term co‑benefits such as improved health, innovation and development \citep{oecd2025}. While debt can constrain investment, climate-oriented debt swaps can repurpose it for adaptation finance \citep{hebbale_2023_debtforadaptation, garcia_2015_barbados}. DRM correlates positively with GNI (PPP) and debt, but negatively with FDI, suggesting mobilization for preparedness alongside investor caution—consistent with prior evidence \citep{neise2022effect}. Adaptation shows limited effects, with a single robust link: a negative coefficient with electricity consumption, hinting at efficiency gains \citep{colelli2022increased}. Two unexpected associations emerge. Mitigation correlates positively with adolescent fertility, diverging from expectations that lower fertility supports climate resilience \citep{hassan_2022_why}, possibly reflecting reverse causality. Similarly, secondary education enrollment shows a negative link with mitigation, echoing \cite{Bragoudakis_2025} that education boosts long-term engagement but may initially raise energy demand. Loss \& Damage exhibits no significant associations, consistent with its limited implementation globally.

\section{Conclusion}

This data-driven, text-based cross-national NLP–econometric analysis reveals significant (and at times unexpected) links between climate policies and development outcomes. Modeling all four policy types jointly reflects their real-world interdependence and enhances policy relevance, highlighting both areas of strong alignment, such as the positive economic impacts of Mitigation, and persistent gaps, notably in Loss and Damage implementation. Beyond its methodological contribution, the framework provides a practical pathway to impact by enabling systematic cross-country tracking, supporting evidence-based climate governance, and informing progress toward the Paris Agreement and Sustainable Development Goals. Future research should explore interactions and non-linearities to better capture the complexity of climate–development relationships and further refine quantitative policy evaluation.

\begin{ack}
The author gratefully acknowledges the support and insights of Arun Kumar Choudhary, Research Scientist at the Ministry of New and Renewable Energy (MNRE), Government of India, provided as part of the Climate Change AI Mentorship Programme. Appreciation is also extended to Prof. Niheer Dasandi (and Prof. Slava Jankin) from the University of Birmingham for proposing this research task as part of the author’s postdoctoral fellowship interview process for the ‘The Lancet Countdown: Tracking Progress on Health and Climate Change’ project, engaging in valuable discussions during and after the process, and encouraging the submission of this work to the workshop.

No funding was received to support this research, and the author declares no conflicts of interest.
\end{ack}

\bibliography{references}
\bibliographystyle{unsrt}

\appendix

\section{About the Data}

\subsection{Climate Change Laws of the World (CCLW) database}
The CCLW database \citep{ccl} is a publicly accessible, comprehensive collection of national climate change legislation and policies from 196 countries and territories, maintained by the Grantham Research Institute and Sabin Center. It tracks laws, regulations, and policy statements related to both climate mitigation and adaptation, including those addressing the transition to a low-carbon economy. As per their website, climate change related policies laws and policies are broadly defined as legal documents that are directly relevant to climate change themes, as discussed in the main text. While the database covers all UNFCCC parties and several territories that are not in the UN or UNFCCC, such as Taiwan, Palestine and Western Sahara, to be included in the database, one or more aspects of a law and policy must have been motivated by climate change concerns. At the time of this manuscript, the dataset included legislation and policy at the national and sectoral levels only, and the documents contain full legal force. Table \ref{tab:topics_cclw} shows the topics that were categorized in the dataset according to the climate policy response. We covered around 1,400 indicators across 214 economies (i.e., a geographic or political unit used for economic measurement, which may or may not be a sovereign country, such as EU). The data contain information curated annually. While many series are available since 1960s, coverage varies by country and variable. Therefore, for simplicity and aligning better with the Paris Agreement era, we analyze the data document onward from the year 2015.

\begin{table}[htbp]
    \centering
    \footnotesize
    \begin{tabular}{|p{2cm}|p{10cm}|}
    \hline
        \textbf{Topic} & \textbf{Meaning} \\
        \hline
        Mitigation & {Mitigation laws and policies are legislative or executive measures aimed at reducing greenhouse gas emissions, either directly (such as through carbon budgets or cap-and-trade systems) or indirectly, by supporting relevant institutions or low-carbon research. Forest and land-use policies are included only if they explicitly contribute to emission reductions or carbon removal; general conservation laws are excluded unless they specifically reference climate change mitigation.}\\ 
        \hline
        Adaptation & {Adaptation laws and policies explicitly address climate change adaptation, requiring adjustments in ecological, social, or economic systems to respond to current or anticipated impacts. A comprehensive review was conducted in 2018, with subsequent additions. These measures are often embedded within broader policies (such as development, planning, disaster management, water, land use, and health) making them sometimes difficult to identify.}\\
        \hline
        Disaster risk management & {Laws and policies on disaster risk management (DRM) and disaster risk reduction (DRR) were added to the database in 2019. Inclusion considers whether they address disasters likely to increase due to climate change (such as hurricanes, floods, heatwaves, droughts, forest fires, or sea-level rise) even if not explicitly climate-motivated, as these laws often take a holistic approach to natural and human hazards.} \\
        \hline
        Loss and damage & {Loss and damage-related laws and policies explicitly aim to reduce climate-related risks by enhancing resilience or providing support to affected individuals and communities. They cover both economic and non-economic harms worsened by human-induced climate change and include measures such as relief funds, insurance schemes, relocation programs, cross-departmental integration, and social protection. The database captures all documents explicitly referencing “loss and damage” since 2015; policies addressing specific climate impacts without this framing are not included.} \\
        
    \hline
    \end{tabular}
    \caption{\small Definitions of the policy categories, as classified in the CCLW dataset}
    \label{tab:topics_cclw}
\end{table}

\subsection{World Bank World Development Indicators (WDI)}
The World Development Indicators (WDI) dataset \citep{worldbank_wdi} is the World Bank's flagship global database of development statistics. It provides internationally harmonized, country-level macro-economic, social and environmental indicators drawn from national statistical agencies, multilateral development institutions, and specialized UN agencies. WDI serves as a benchmark source for cross-country development monitoring and policy evaluation across academic, institutional and policy domains. 
WDI indicators are used as macro-level outcomes reflective of national development pathways; they do not capture sub-national distributional effects or project-level policy impacts.
Table \ref{tab:wdi_data_notes} reports the data notes for the dataset used in our study.

\begin{table}[htbp!]
    \centering
    \footnotesize
    \begin{tabular}{|p{2cm}|p{10cm}|}
    \hline
       Aspect & Details \\
       \hline
        Coverage & 2015-present (aligned with Paris Agreement implementation period)\\
        Countries & $\approx$180 countries (subjected to indicator availability)\\
        Variables used & {GDP (current USD), GNI (Atlas method \& PPP), FDI inflows, external debt stocks, electricity consumption, adolescent fertility rate, secondary school enrollment}\\
        Transformations & {Standardized, logged where applicable, country \& year fixed effects applied} \\
    \hline
    \end{tabular}
    \caption{WDI data notes}
    \label{tab:wdi_data_notes}
\end{table}

\paragraph{Why use this data?}
The WDI dataset is well suited in this analysis because it provides a harmonized, globally comprehensive set of macro-economic and social indicators that enable consistent cross-country comparison over time. As the World Bank's flagship development database, WDI is widely used in empirical research and policy evaluation by institutions including UN, IMF and IPCC-aligned initiatives, making it appropriate for linking national-level climate policy emphasis to socio-economic contexts. The indicators selected: spanning economic capacity (GDP, GNI), financial conditions (FDI, external debt), infrastructure and energy use (electricity consumption), and demographic or human capital factors (fertility, education); capture structural characteristics that shape countries’ ability to plan and implement climate policy. Because WDI reflects official statistics reported through standardized international protocols, it offers both reliability and comparability across regions and income groups. While national-level aggregates cannot capture local variation in exposure, implementation, or equity outcomes, WDI provides a rigorous foundation for cross-national analyses of climate–development relationships and serves as an appropriate first step toward deeper causal and subnational inquiry.

\section{Data Preparation for Statistical Analysis}

\subsection{Data Harmonization and Processing}
WDI indicators were harmonized into a consistent panel at the country-year level for 2015 onward to align with the post-Paris Agreement period of policy reporting. Where applicable, variables with skewed distributions (e.g., GDP, GNI, FDI, debt) were natural-log transformed to improve interpretability and reduce the influence of outliers. The analysis uses an unbalanced panel structure, retaining all observed values and avoiding imputation so as not to introduce modeling assumptions into the macro‐level data. Indicator units and country identifiers were standardized, and observations were matched by ISO3 codes and calendar year. To ensure comparability across indicators, nominal values were converted where necessary using WDI’s harmonized methodology (e.g., Atlas method and PPP for GNI), and each indicator’s temporal alignment was inspected to avoid structural breaks or definitional changes. No smoothing or interpolation was applied. These steps preserve the fidelity of official statistics while enabling reproducible cross-country econometric analysis.

\subsection{Data Merging}
The decision to merge the climate‑policy dataset (CCLW) with the socioeconomic dataset (WDI) was driven by the analytical goal of assessing how national economic conditions and development contexts relate to countries’ climate‑policy orientations. By combining policy‐measure indicators (from CCLW) with metrics of development, income, and structural characteristics (from WDI), the resulting dataset enables multivariate statistical modelling to explore associations, clusters, and variance in policy emphasis across countries. In practical terms, the join allows each country to appear in the same observation row with both its climate legislation profile and its socioeconomic attributes, thereby facilitating techniques such as principal component analysis, biplots, and regression modelling--- all of which was used in this exploratory statistical analysis. Without this merge, analyses would have been limited to either the policy side or the economic side alone; the integrated dataset is, therefore, essential for identifying how development capacity, income level and structural factors correlate with the emphasis on different climate‑policy domains.

\section{Descriptive Policy Patterns}
\begin{figure}[!htbp]
  \centering
  \includegraphics[width=0.7\linewidth,]{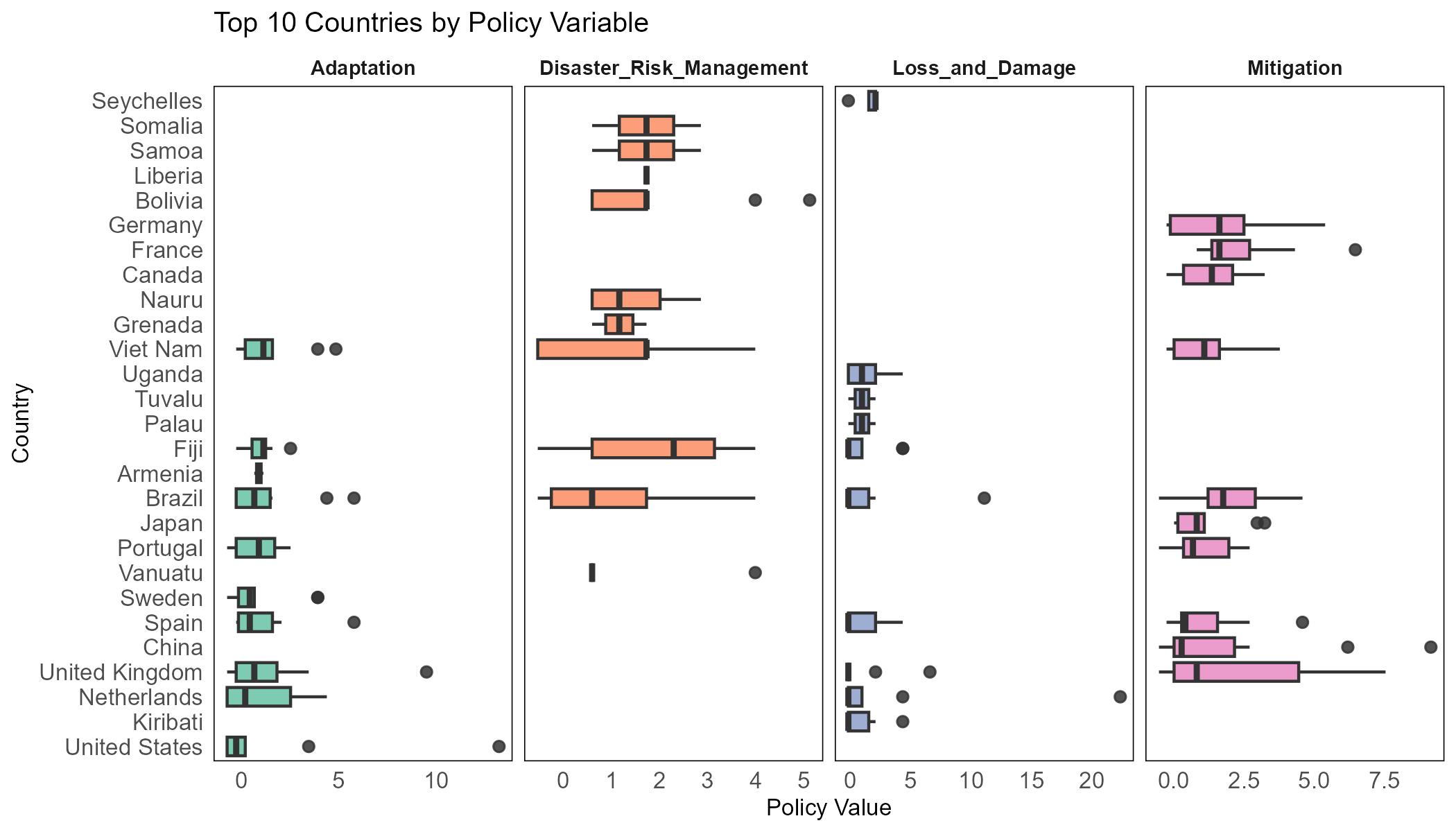}
  \caption{\small This figure shows faceted boxplots of the top 10 countries for each policy variable, highlighting comparative differences in policy implementation.}
  \label{fig:boxplot}
\end{figure}

The faceted box plot in Figure \ref{fig:boxplot} shows standardized thematic policy emphasis (z-scores) across the top decile of countries for each theme. While consistent with known capacity and vulnerability patterns in climate governance, the plot is included here for completeness, as the main narrative focuses on cross-national structure (via correspondence analysis) and links to development outcomes (via fixed-effects regression).

The figure illustrates climate policy engagement across the top ten countries for Adaptation, Disaster Risk Management (DRM), Loss and Damage, and Mitigation. Data were reshaped into long format and standardized (z-scores), with top performers selected by relative intensity rather than raw counts. Box widths reflect temporal variation from 2015 onward, while higher absolute z-scores denote stronger policy emphasis. Mitigation dominates, led by developed nations (Germany, France, Canada), whereas Adaptation and DRM are driven by climate-vulnerable small island developing states (SIDS). Loss and Damage shows minimal global activity, indicating underdevelopment. These contrasts underscore how national circumstances shape priorities: developed economies emphasize mitigation, vulnerable nations focus on adaptation and risk management. 

\section{Limitations and Directions for Future Work}
This study has several limitations. First, the statistical analysis identifies associations, not causal effects. Although two-way fixed effects reduce bias from time-invariant country characteristics and global shocks, the estimates may still reflect omitted variable bias, reverse causality, and policy endogeneity. Climate policy intensity can both influence and be influenced by development variables (e.g., wealth, debt capacity, education expansion, demographic trends), limiting causal interpretation. Second, policy text signals may not fully capture implementation quality or budgetary effort, potentially understating “on-the-ground” adaptation and DRM actions. Third, class imbalance (particularly for Loss \& Damage) limits predictive granularity for emerging policy domains. Finally, the time horizon is constrained to 2015 onward due to the Paris Agreement reporting period and availability of harmonized climate policy text data.
As the future work, we will be extending this analysis using quasi-experimental designs, lag structures, instrumental variables, and granular budget/implementation data to better estimate causal climate-development pathways.

\end{document}